\documentclass[preprint,aps,floatfix,groupedaddress,showpacs,showkeys,preprintnumbers,amsmath,amssymb]{revtex4}
\usepackage{graphicx}
\begin{document} 

\title{Nuclear condensation and the equation of state of nuclear matter}

\author{J. N. De and S. K. Samaddar}
\affiliation{
Saha Institute of Nuclear Physics, 1/AF Bidhannagar, Kolkata
{\sl 700064}, India} 


\begin{abstract}
The isothermal compression of a dilute nucleonic gas invoking
cluster degrees of freedom is studied in an equilibrium statistical
model; this clusterized system is found to be more stable than 
the pure nucleonic system. The equation of state (EoS) of this matter,
shows features qualitatively very similar to the one obtained from
pure nucleonic gas. In the isothermal compression process, there
is a sudden enhancement of clusterization at a transition density
rendering features analogous to the gas-liquid phase transition in normal
dilute nucleonic matter. Different observables like the caloric curves,
heat capacities, isospin distillation, etc are studied in both
the models. Possible changes in the observables due to recently
indicated medium modifications in the symmetry energy are also
investigated. 
\end{abstract} 



\maketitle

\section{Introduction}

 Nuclei expand with excitation. If the excitation energy per 
nucleon $E^*/A$ is sufficiently high, the compound nuclear 
configuration is no longer a stable one and the hot nuclei
disassemble into many fragments. This general observation
\cite{koo,bon,gro} from intermediate energy heavy ion collisions 
has been termed as nuclear multifragmentation. Beyond a 
certain excitation, the nuclear system is also found to
vaporize mostly into nucleon gas \cite{riv,bor}  with all produced
fragments having atomic numbers $<$ 3.
The associated
temperature corresponding to this excitation can be identified as
the boiling temperature of the fragmenting nucleus.

The study of the correlation of temperature with the total
excitation energy obtained from the energetics of the generated
fragments generally reveals a plateau \cite{poc,cib} over an excitation
energy range of around 3 MeV $\leq E^*/A \leq $ 10 MeV. 
This is reminiscent of the liquid to gas phase
transition in a thermodynamic system, 
understood as a crossover of the denser homogeneous
nuclear matter to a continuum of subnuclear densities where
nucleons and larger nucleonic clusters coexist in thermal and
chemical equilibrium. This basic picture is at the heart of
various statistical models \cite{koo,bon1,gro} for multifragmentation
that have been quite successful in explaining many features of
the experimental data.

The laboratory observables on nuclear multifragmentation
offer a platform for understanding complex processes in
astrophysical context that depend on properties of low density
nuclear matter, like core-collapse supernovae, giant stellar
explosions or element formation in explosive nucleosynthesis.
Shredded of the minute details, it would be interesting to
know what happens to a dilute hot gas of interacting nucleons when
compressed isothermally or cooled isochorically? Answers have generally
been sought from the statistical mechanics of such an interacting
system. Detailed calculations in the relativistic mean-field
framework \cite{mul} have been done for the EoS 
of infinite nuclear matter of different neutron-proton
concentrations. The main conclusions are: i) the nucleons in the 
gas phase condense to a bulk liquid phase with which it remains
in coexistence over a range of densities, ii) at constant pressure
the gas-liquid phase transition occurs at a constant temperature
for the symmetric system ($N=Z$) alluding to a first order
phase transition; for the asymmetric systems, however, isobaric
transition occurs over a range of temperatures indicating 
a continuous transition and iii) the neutron-proton asymmetry
in the gas and liquid phases are in general widely different
for asymmetric matter (isospin distillation). Similar conclusions 
have also been arrived at \cite{sil} in a nonrelativistic 
mean-field model.

The mean-field calculations do not account for fluctuations in the
nucleonic degrees of freedom that render it possible for cluster
formation from the nucleonic gas. On that account, the method of
virial expansion has been found to be quite useful and practical
in calculating the EoS of dilute gases \cite{hua}. The formal
structure of the virial coefficients are self-contained,
nonetheless, given an interaction potential, the calculation of higher
virials are rather tedious. In quantum statistical mechanics, the second
virial coefficient may be expressed in terms of scattering phase shifts. 
Using the modern language of diagrammatics, the general discussion on
virial coefficients becomes very elegant and concise \cite{das,hor}, but
the fact remains that a practical calculation of the virial coefficients
beyond the second is an onerous task. 

 The explicit evaluation of the EoS of the interacting dilute
gas from the classical virial (cluster) expansion is very cumbersome,
but one thing that transpires clearly from the method is
that the interacting gas can be treated as a noninteracting
gas of clusters of different sizes \cite{pat} in thermodynamic equilibrium.
In this backdrop, we study the nuclear EoS of both symmetric
and asymmetric matter exploiting the nuclear statistical
equilibrium (NSE) model as has been employed in understanding the 
nuclear multifragmentation data or in exploring nucleosynthesis
in the astrophysical scenario \cite{mey,ish,bot}. Explicit consideration
of the interaction between nucleons as is required in the
cluster expansion is bypassed in this model; their effect is
indirectly borne through the binding energies of the clusters,
which are taken as phenomenological inputs. 

    Recent laboratory experiments \cite{lef,igl} on nuclear 
disassembly indicate that the properties of the nuclides
describing their binding energies are modified at the 
subnuclear densities ($\rho \sim \rho_0/3$) they are created
in. The symmetry energy, for example, is reported \cite{she}
to be progressively reduced with excitation energy which is
attributed to the in-medium modifications of the properties
of the hot fragments \cite{sou} or their expansion \cite{sam,she1}.
Similarly, the surface properties of the hot fragments are speculated
to be modified due to the embedding environment \cite{bot1}. The
highest density that we explore in our calculation is relatively
dilute compared to the freeze-out density in which the fragments are
formed in laboratory experiments. We therefore do not expect
the in-medium modification of the surface energy to play any
significant role in the present context. However, modifications 
 in the symmetry energy are considered as they may arise from
expansion of the hot fragments.

    The paper is organized as follows. In Sec. II, we
briefly outline the statistical equilibrium model. The results and discussions are
presented in Sec. III and the concluding remarks
are given in Sec. IV.

\section{The statistical equilibrium model}
 
  We work in the framework of the grand canonical ensemble. Taking the 
cue from Ursell and Meyer \cite{pat}, we assume the interacting dilute
nucleon gas to be a noninteracting mixture of 
nucleons and different nucleonic
clusters in thermal and chemical equilibrium.
For simplicity of terminology, we henceforth term all 
the species as particles or fragments that include
monomers (neutrons and protons) and more complex
clusters.  The neutron and proton
densities are conserved on the average as
\begin{eqnarray}
\sum_i N_i \rho_i = \rho_n, \nonumber \\ 
\sum_i Z_i \rho_i = \rho_p, 
\end{eqnarray}
where $\rho_n$ and $\rho_p$ are the total neutron and proton densities of 
the nuclear matter and $\rho_i$ is the number density of the $i^{\rm th}$
fragment species with $N_i$ neutrons and $Z_i$ protons. 
The sum in Eq.(1) extends over all the cluster species. 
The fragment number density
$\rho_i$ is obtained as
\begin{eqnarray}
\rho_i = \frac{g_i}{h^3}\int n_i({\bf p_i})~ d{\bf p_i}.
\end{eqnarray}
Here $g_i$ is the degeneracy factor, ${\bf p_i}$ refers to the 
momentum of the $i^{\rm th}$ fragment species and $n_i$ is the distribution
function given by
\begin{eqnarray}
n_i({\bf p_i}) = \left [exp(\varepsilon_i-\mu_i)/T \pm 1
\right ]^{-1},
\end{eqnarray}
where
\begin{eqnarray}
\varepsilon_i = \frac{p_i^2}{2m_i}-B_i,
\end{eqnarray}
is the single particle energy of the fragment species, $m_i$
its mass, $B_i$ the ground state binding energy
and $T$ is the temperature of the system. From the 
condition of chemical equilibrium, one gets for chemical
potential $\mu_i$ of the species,
\begin{eqnarray}
\mu_i = N_i \mu_n+Z_i \mu_p,
\end{eqnarray}
where $\mu_n$ and $\mu_p$ are the chemical potentials of the
monomers in chemical equilibrium with
the clusters.  These are determined iteratively from
the conservation conditions given by Eq.(1). The +($-$) signs in
Eq.(3) refer to fermions (bosons). The density of the fragment
species is obtained after momentum integration of the occupation
function (Eq.(2)); for fermions, 
taking into account the various excited states of the fragment
species, it is given by
\begin{eqnarray}
\rho_i = \frac{2}{\sqrt{\pi}\lambda ^3} A_i^{3/2}
J_{1/2}^{(+)}(\eta_i) \phi_i(T).
\end{eqnarray}
Here $\phi_i(T) $ is the internal partition function
\begin{eqnarray}
\phi_i(T) & = & \sum_k g_i^k e^{-E_i^k/T} \nonumber \\
& = & e^{-F_i/T},
\end{eqnarray}
where $E_i^k$ is the internal excitation energy of the $i^ {\rm th}$
species for the $k^ {\rm th}$ state and $F_i$ is the internal free
energy of the cluster measured with respect to its ground state.
Fragments with mass $\leq$ 4 are assumed to have no internal
excitation.
Similarly, the density of the bosonic fragments is
\begin{eqnarray}
\rho_i = \frac{g_i^0}{V} \frac{1}{exp(-\eta_i)-1}+
 \frac{2}{\sqrt{\pi}\lambda ^3} A_i^{3/2}
J_{1/2}^{(-)}(\eta_i) \phi_i(T).
\end{eqnarray}
In Eqs. (6) and (8), $\lambda =\frac{h}{\sqrt{2\pi mT}} $ is the
thermal wavelength of a nucleon ($m$ is the nucleon mass,  
we assume the neutron and proton mass
to be the same), $\eta_i = (\mu_i+B_i)/T$, $A_i$ ($=N_i+Z_i$) is
the fragment mass number, and $\phi_i (T)$ is the internal partition
function of the excited fragments which reduces to the ground state
degeneracy factor $g_i^0$ in the limit $T \rightarrow 0$. The functions 
$J_n^{(\pm )}$ are the Fermi (Bose) integrals and $V$ is the volume
of the system. The explicit form of the integrals is given by
\begin{eqnarray}
 J_n^{(\pm )}(\eta ) = \int_0^\infty \frac{x^n}{exp (x-\eta )
\pm 1}~dx.
\end{eqnarray}
The first term in Eq.(8) arises from bose condensation which
we neglect in subsequent calculations as it turns out to be 
very insignificant in the present context. 
For the ground state binding energy of the fragments, we take recourse
to the recently proposed mass formula by Danielewicz \cite{dan}.
We have also used the Myers-Swiatecki \cite{mye1} mass formula,
but many nuclear EoS properties investigated with both mass 
formulas look indistinguishable. We therefore report calculations 
with the Danielewicz mass formula. In this model, the binding energy
of a chargeless nucleus is,
\begin{eqnarray}
B_i = a_vA_i~-\sigma (X_i,T=0)A_i^{2/3}~-\alpha 
\left (\frac{N_i-Z_i}{A_i} \right )^2A_i +\delta .
\end{eqnarray}
For monomers, $B_i$ is zero.
Since the fragments are formed out of chargeless infinite nuclear matter,
the coulomb term has been dropped. In Eq.(10), the first term is the 
volume term with volume energy coefficient $a_v$=15.6163 MeV,
$\sigma (X,T)$ is the surface energy coefficient and $\alpha$
is the  volume symmetry energy coefficient, taken as 32.6655 MeV.
The temperature and asymmetry ($X=~(N-Z)/A$) 
dependent surface energy coefficient is \cite{lev}
\begin{eqnarray}
\sigma (X,T) = \left [\sigma (X=0,T=0)-a_sX^2 \right ]
\left [1+\frac{3}{2}\frac{T}{T_c} \right ]
\left [1-\frac{T}{T_c} \right ]^{3/2},
\end{eqnarray}
with $\sigma (0,0)$ =17.9878 MeV. The surface symmetry energy coefficient
$a_s$ is given by $a_s=(\alpha^2/\beta)/(1+\alpha /\beta 
A_i^{-1/3})$ with $\beta $=13.6106 MeV.  
The critical temperature of
symmetric nuclear matter $T_c$ is taken to be 14.61 MeV. It
pertains to the
SkM$^*$ effective interaction \cite{kol} which we use in subsequent
calculations. The last term in Eq. (10) is due to pairing which
is zero for odd-even nuclides and is equal to $\delta = \pm
a_p/\sqrt {A_i}$ for even-even or odd-odd nuclei.
For ground state, the value of
$a_p$ =10.8714 MeV; as pairing vanishes at above $T \sim $1 MeV, it is
taken to be zero for the NSE calculations reported here.
For fragment mass $A_i$ $>$ 4, Eq.(10) is employed
for calculation of binding energies. For fragments with 
$A_i \leq $4, they are calculated by subtracting the coulomb
contribution from the experimental binding energies. 

The free energy $F_i$ is obtained in the Fermi gas approximation
where the volume excitation energy is $a_iT^2$ and the volume entropy
$S_i^{vol}$ is $2a_iT$ with $a_i$, the level density parameter
of the $i^ {\rm th}$ species being taken as $a_i=A_i/16$. Then 
we have,
\begin{eqnarray}
F_i= \left [\sigma (X_i,T)-\sigma (X_i,T=0) \right ]A_i^{2/3}
-a_iT^2,
\end{eqnarray}
where the first term is the excess surface free energy due to 
excitation and is designated as $F_i^{sur\!f}$. The excitation
energy per nucleon of the disassembled system is then given
by,
\begin{eqnarray}
E^*/A= \left (\sum_i \rho_i E_i(T) + \frac{3}{2} \sum_i
\rho_i T \right )/\sum_i \rho_i A_i,
\end{eqnarray}
where the internal excitation $E_i(T)$ of the $i^ {\rm th} $ species
is
\begin{eqnarray}
E_i(T)= F_i+T\left ( S_i^{vol}+S_i^{sur\!f} \right ).
\end{eqnarray}
The second term in the numerator of Eq.(13) is the kinetic
energy density for the center of mass motion of the particles.
Here we have used the classical equipartition theorem which is
a very good approximation to the quantal result for the 
dilute system.
The surface entropy $S_i^{sur\!f}$ is obtained from 
$S_i^{sur\!f}= -\partial F_i^{sur\!f}/\partial T$. In addition to the
surface and volume terms, the total entropy has a contribution
$S_i^{tran}$ from the translational motion of the center of mass
of the fragments. The total entropy per nucleon generated from 
the disassembled nuclear matter is then
\begin{eqnarray}
S/A~=~\sum_i \rho_i \left (S_i^{sur\!f}+S_i^{vol}+S_i^{tran}
\right )/\sum_i \rho_i A_i,
\end{eqnarray}
where the translational entropy is taken as \cite{bon2}
\begin{eqnarray}
S_i^{tran}= \frac{5}{2} +ln \left (g_i \frac{A_i^{3/2}}
{\rho_i \lambda ^3} \right ).
\end{eqnarray}
In Eq.(16), for the degeneracy factor $g_i$, the experimental
ground state degeneracy is taken for light fragments
($A \leq 16$), otherwise, it is taken to be two for fermions 
and one for bosons. 

The pressure is calculated from the total free energy $F$
as $P=-\partial F/\partial V$ which comes out as
\begin{eqnarray}
P= \sum_i \rho_i T,
\end{eqnarray}
the sum of the partial pressures of a mixture of noninteracting
particles consisting of clusters and monomers. The total
multiplicity is $M=\sum_i\rho_iV$.

The quantum distribution function  approaches
the classical one when $\eta_i= (\mu_i+B_i)/T$ $<<$~0. In that
case, $J_{1/2}^{(\pm )}(\eta )\rightarrow \frac{\sqrt \pi}{2}
e^{\eta }$ and then
\begin{eqnarray}
 \rho_i = \frac{A_i^{3/2}}{\lambda ^3}e^{\eta_i }\phi_i(T).
\end{eqnarray}
In the range of densities and temperatures we explore in
the present work, this condition is mostly satisfied. Examination
of Eqs.~ (17) and (18) shows that the pressure 
and the density can be written in the
form,
\begin{eqnarray}
P= T\sum_i \frac{1}{\lambda^3}d_i (z_n)^{N_i} (z_p)^{Z_i},
\end{eqnarray}
\begin{eqnarray}
\rho & = & \sum_i~ \rho_i A_i \nonumber \\
& = & \sum_i\frac{A_i}{\lambda^3}d_i(z_n)^{N_i}(z_p)^{Z_i},
\end{eqnarray}
with  fugacity $z_{n,p}=e^{\mu_{n,p}/T}$ and 
$d_i=A_i^{3/2} e^{B_i/T}
\phi_i(T)$. For a one component system, one readily sees that
the above equations have the structure analogous to the ones obtained
from cluster expansion \cite{hua}
\begin{eqnarray}
P/T=\frac{1}{\lambda^3} \sum_i b_iz^i,
\end{eqnarray}
\begin{eqnarray}
\rho =\frac{1}{\lambda^3} \sum_i ib_i z^i,
\end{eqnarray}
with  $z=e^{\mu /T}$ ($\mu_n =\mu_p =\mu$)and $b_i$ is the cluster integral.
In cluster expansion, information on the two-body interaction is
contained in the cluster integral $b_i$, similarly in the NSE
model, this information is embedded as binding energy in $d_i$. 
The two models are not equivalent; an $i-$particle cluster in
the cluster expansion includes both bound and continuum states
and the $b_i$'s in this expansion can be negative \cite{pat}.
The $d_i$'s in the NSE model are, however, always positive,
they include only a variety of bound states.

\section{Results and discussions}

In the mean-field models of gas-liquid phase transition \cite{mul,sil},
the hot dilute nuclear gas is assumed to consist only of monomers.
With isothermal compression or isochoric cooling, a critical stage is 
reached when there is a sudden onset of bulk liquid formation in
coexistence with the monomers. This is the condensation point
(the gas-liquid transition). We refer to this model as NM model.
In the NSE model, even at very low density, in addition to monomers,
the system may contain
dimers and lighter clusters, though in macroscopically very 
insignificant amount. With compression or
cooling, the clusters grow in size which becomes critical around
the condensation point. Thermodynamically, this is more favorable
compared to the pure monomeric picture. This is evident from a comparison
of the free energy per nucleon which is displayed in Fig.~1, calculated at
a temperature $T$=~5 MeV, with both the NSE and NM models, for
symmetric and asymmetric nuclear matter. Details of the calculations
in the NSE model are postponed till the next paragraph. 
It is seen that at very
low density when cluster formation is insignificant, the two
free energies are indistinguishable. We restrict this calculation
in a relatively dilute regime where interactions in the NM 
model can be ignored.

In infinite matter, the maximum cluster size $A_f^{max}$ can,
in principle, be infinite. Numerical calculations necessitate
the restriction of $A_f^{max}$ to a finite number. The 
thermodynamic observables are sensitive  to the choice of
$A_f^{max}$. The results, however, tend to converge with
increasing $A_f^{max}$ which is displayed in Fig.~2 for
isotherms drawn at $T$=~5 MeV for nuclear matter of different
asymmetries ($X_t$). In panels $(a), (b)$ and $(c)$, the isotherms 
are drawn for asymmetries $X_t$= 0.0, 0.2 and 0.4, respectively
for different $A_f^{max}$ and it seen that they tend to converge
at $A_f^{max}$ $\sim$ ~5000, the convergence being faster with
increasing asymmetry. The results in Fig.~1 and 
all the subsequent calculations reported 
are performed with $A_f^{max}$ =5000
unless otherwise mentioned. The two-component nature
of the system allows formation of asymmetric fragments for even
symmetric nuclear matter ($X_t$=0.0). 
With the neglect of the difference between the neutron and proton
masses, all the observables in symmetric nuclear matter are 
invariant with respect to interchange of neutron and proton
and hence the system is effectively a one-component system.
Absence of coulomb interaction
helps in formation of fragments with lower asymmetry and therefore
consideration of fragments with asymmetry in the range
$-$0.3 $\leq X_i \leq $ 0.3 suffices, even for very asymmetric
nuclear matter.
To facilitate the comparison of the dependence of the isotherms
on the asymmetry of nuclear matter in the NSE model, they are
shown in panel $(d)$ of the figure at $T$=5 MeV, for $X_t$
=0.0, 0.2 and 0.4. Unlike the NM model, a sharp condensation
point can not be defined in the NSE model, however, a break in
the $P- \rho$ curve is apparent when larger clusters start 
to form suddenly, the break being sharper with lesser asymmetry.
In this panel, calculations with the Myers-Swiatecki mass
formula are also displayed; they are practically identical
with the ones performed with the Danielewicz mass formula.
The evolution of the isotherms with asymmetry in the NSE model
is very similar to those \cite{mul} obtained from the NM
model; for symmetric nuclear matter, the condensation occurs
practically at constant pressure whereas for the asymmetric
systems, the pressure increases. Such a behavior with clusterization
has also been reported earlier \cite{das1}. 
The chemical equilibrium conditions
coupled with the conservation of the nucleon number and
isospin underlies this behavior. 

    On a somewhat formal footing, this is understood by exploiting
Eq.(20). For simplicity, we consider a one-component system,
extension to two-component system is straightforward. The
population of a cluster of size $i$ in the one-component
system can be written as
\begin{eqnarray}
\omega_i = \frac{V}{\lambda ^3}c_i (b_0z)^i,
\end{eqnarray}
with $c_i=i^{3/2}exp\left (-\sigma (T)i^{2/3}\right )$ and $b_0
=exp\left (a_v+\frac{T^2}{16} \right )$. As the $i -$dependence of $c_i$
is relatively weak compared to $(b_0z)^i$, when $i >>1$ and
$z<b_0^{-1}$, $\omega_i$ is practically
zero and one can consider condensation insignificant in a 
macroscopic sense. As $z$ increases and passes through the critical
value, the density given by the sum in Eq.(22) shoots up very fast.
In other words, the density is a very sensitive function of $z$
or the chemical potential, which thus shows a near-constancy
in the condensation region. The pressure is an explicit function
of $z$ and $T$ and depends on the density only in a thermodynamically
negligible manner and thus shows the same features as the 
chemical potential. For asymmetric systems, say,  with neutron 
excess, the population of clusters with $N_i$ neutrons and
$Z_i$ protons is

\begin{eqnarray}
\omega_i = \frac{V}{\lambda ^3}c_i (b_{0,n} z_n)^{N_i}
(b_{0,p} z_p)^{Z_i}.
\end{eqnarray}
Formation of symmetric fragments is more probable from binding
energy considerations, hence as larger fragments start to
form, free neutrons become more in excess over protons 
(isospin distillation) and the $z_n$ starts to increase 
and $z_p$ decreases. This extra degree of freedom keeps
the situation more subtle in asymmetric matter resulting in
increase of pressure with density in the transition region.

In Fig.~3, the nucleon fraction constituting the gas phase or the 
liquid phase as a function of nuclear density is displayed at 
$T$= 2 and 5 MeV for symmetric nuclear matter. The monomers are
considered as gas while the rest (dimers and larger clusters)
constitute the liquid part. The baryon fraction
$R$ is defined as $R_{l,g}= A_{l,g}/(A_l+A_g)$ where $l,~g$
stand for the liquid and gas phases, $A_{l,g}$
being the nucleon number in the respective phases.
With increasing density, the liquid fraction $R_l$ initially
rises very fast and then gradually approaches unity. The
behavior of the gas fraction is just the opposite. In 
panels $(a)$ and $(b)$ of this figure, the effect of the 
choice of $A_f^{max}$ at the two temperatures is shown.
It is seen that the choice of the
maximum cluster size we made for our calculations
is satisfactory. In panels $(c)$ and
$(d)$, the liquid and gas fractions in the NSE and the
NM models are compared at the two temperatures. They are not too
different; at high densities, these fractions tend to
merge in these model calculations. 

In the upper panel of Fig.~4, $A_l/A_g$, the ratio of 
the numbers of nucleons in  
the liquid to gas phase as a function
of density for asymmetric nuclear matter ($X_t$=0.4)
in the NSE and the NM models at $T$=5 MeV are compared.
As with symmetric nuclear
matter, here also there is not much significant difference
in the results from the two models.
The intersection of the dashed line with the abscissa
defines the condensation point in the NM model below
which there is no liquid; in the NSE model, however, the
liquid phase starts at a lower density as there can still be 
some light clusters in the dilute matter. The lower panel
in the figure displays the isotherms for the said system.
The overall behavior of the pressure with density in both
the models are the same, the pressure increases with density
beyond the condensation point as opposed to symmetric
nuclear matter. The pressure is seen to be somewhat
lower in the NM model. This is understandable.  
In the NM model, the pressure in the coexistence region is
given either by the liquid pressure or the gas pressure
($P=P_l=P_g$) whereas in the NSE model, the sum of the
two pressures ($P=P_l+P_g$) has to be counted.

 In Fig.~5, the response of the dilute nuclear systems of 
different neutron and proton concentration to cluster formation
on isochoric cooling is displayed. The upper and lower panels
correspond to densities $\rho =$ 0.0002 and 0.002 fm$^{-3}$.
The rate of change of the total multiplicity $M$ per baryon 
with temperature ($A^{-1}dM/dT$, $A$ refers to the total number
of nucleons in the system) is taken to be a measure of this
response. At higher $T$, the system is mostly in the monomeric
phase. At a particular density, as the system cools down, 
clusters start to form and at a particular temperature, 
there is a sudden growth of cluster formation. This temperature
is the condensation temperature which we refer to as the boiling temperature
$T_b$ corresponding to this density. As the system is cooled further,
the multiplicity decreases because of the formation of larger
clusters at the cost of smaller ones which is reflected in the
reduction of $dM/dT$. The boiling temperature is found to be
rather insensitive to the neutron-proton asymmetry, there is a  
nominal decrease in $T_b$ with $X_t$. The decrease in boiling temperature
with decreasing density is more marked. 

For dilute matter, the total multiplicity $M$ (=$A$ in the limit of
large specific volume $v=1/\rho $) decreases with isothermal
compression. This is shown in Fig.~6 where we display $dM/dv$,
the rate of change of multiplicity with specific volume 
as a function of $v$, at a temperature $T=$ 5 MeV for 
different values of $X_t$. The multiplicity $M$ always decreases
in isothermal compression, hence $dM/dv$ is always positive.
The discontinuity in $dM/dv$ indicates the suddenness in
cluster formation at the condensation point. For symmetric
matter, the constancy in $dM/dv$ below condensation volume is
a reflection of the constancy of pressure in the transition
region. For asymmetric matter, similarly, the fall of $dM/dv$
is indicative of the increase of pressure with density
in this region.

In the upper panel of Fig.~7, the caloric curve at a constant
density ($\rho =0.002~ fm^{-3}$) for a representative asymmetric
system with $X_t$ =0.2 is displayed. The full and dash lines
refer to the NSE and the NM models, respectively. Comparison of
the two curves shows that at the same excitation energy, the
temperature in the NSE model is lower because there is some energy
locked up in the creation of surfaces of the clusters. Both the caloric 
curves have a characteristic plateau in temperature; this signals
a liquid-gas type phase transition which is more apparent from the
peaked structure of the 
derivative of the caloric curve, namely, the heat capacity 
per baryon $c_v$
which is displayed in the lower panel of the figure.

The caloric curves for asymmetric nuclear matter ($X_t=0.2$) at
constant pressures of $P$ =0.002 and 0.02 MeV fm$^{-3}$ are
shown in the upper panel of Fig.~8. In both the NSE and NM
models, it is seen that the temperature remains nearly constant
over a broad excitation energy interval. The corresponding
heat capacity per baryon $c_p$ shows very sharp peaks as can be seen
from the lower panel of the figure. At high pressure, in both
models, the transition temperature is higher; the NSE model
shows a lower transition temperature compared to that in the 
NM model due to reasons as explained in the preceding
paragraph. The corresponding
entropies per nucleon $S/A$ as a function of temperature 
are displayed in Fig.~9. Near the transition temperature,
a sharp rise in the entropies can be noted. It may be pointed
out that for symmetric nuclear matter, the liquid-gas type
phase transition occurs at a constant temperature resulting
in a singularity in $c_p$ and a discontinuity in the entropy.

Symmetric cluster formation is more favorable from binding
energy considerations. In asymmetric, say, neutron-rich
nuclear matter, the monomers left after cluster formation
would therefore be richer in neutrons with increase in density. This 
distillation in isospin is also found in the NM model where from
thermodynamic equilibrium conditions, the liquid in 
the coexistence region tends to become symmetric making  the
gas phase richer in neutrons with increasing liquid phase.
The neutron-proton density ratio $\rho_n /\rho_p $ in the gas and
liquid (clusters in the NSE model) in both the models are
displayed in Fig.~10. For the liquid phase, this ratio
varies from $\sim$ 1.0 at low density to $\sim$ 1.5 
at higher density (which is the value of the ratio for
asymmetry $X_t$ =0.2) 
and is practically indistinguishable in the two models.  However, in
the gas phase, there is a dramatic increase in the neutron
number over the proton number beyond the transition density; this is
more pronounced in the NSE model. Calculations in the NSE model
with the Myers-Swiatecki mass formula is also shown in the figure.
Though the inclusive observables (like $P-\rho$ correlation)
are nearly indistinguishable, isospin distillation shows marked
differences between the two mass formulas used. Lesser isospin
distillation in the Myers-Swiatecki mass formula is a reflection of
its smaller asymmetry energy coefficient compared to that 
in the Danielewicz formula.

As mentioned in the introduction, analysis of recent experimental
data indicates a progressive reduction of the symmetry energy
coefficient with the excitation energy of the disassembled system.
In the analysis with the SMM multifragmentation model, the
freeze-out density is $\sim \rho_0/3$ (where $\rho_0$ is the
normal nuclear matter density), but the nuclear interaction 
between the produced fragments is neglected which tentatively shows up as
in-medium correction to the symmetry energy. Moreover, the possible
expansion of the hot fragments is likely to reduce the symmetry
energy. In our present context, since we explore from the
very dilute density regime to $\rho \sim \rho_0/8$, we do
not expect the in-medium corrections to be significant, however,
the expansion of the hot fragments might weaken the symmetry
energy. {\it Ab initio} calculations of these effects on symmetry energy
are difficult, however, to have a feel how the reduced symmetry energy
affects the nuclear observables, we have also performed calculations
with symmetry energy reduced by 40$\%$ at $T$=5 MeV. The corresponding
results are displayed in Fig.~11. The full lines correspond to
the regular system ({\it i.e.}, the system with the normal
value of the symmetry energy coefficient)
and the dashed lines to the one with the reduced value
of the symmetry energy in all the panels. In panel (a) of this figure,
the nuclear EoS ($P-\rho $ correlation) for both symmetric and
asymmetric ($X_t=0.2$) nuclear matter are shown. Reduction of
the symmetry energy coefficient increases the binding energy
of the asymmetric fragments and the chemical potential decreases;
the increased binding energy tends to reduce the total multiplicity by
producing heavy fragments whereas the role of the chemical potential
is opposite. A delicate interplay of these two effects is manifested
in a very little increase of pressure with reduction of symmetry
energy coefficient in case of symmetric matter. This interplay,
however, has a different role in case of nuclear matter with
sizeable asymmetry. For asymmetric matter with reduction in the
symmetry energy, more neutrons can be accommodated in relatively heavier 
fragments and the neutron gas is depleted with reduction of the
total multiplicity and hence the pressure. Inspection shows that the
results for an asymmetric matter with a reduced symmetry 
energy coefficient are similar to
those of regular system of effectively lower asymmetry. 

In panel (b), the neutron-proton density ratio in the liquid
and gas phase as a function of density for $X_t$=0.2 is displayed for
both normal and reduced symmetry energies. In the liquid phase,
the two results are very close; in the gas phase, as already 
mentioned, the neutron multiplicity gets depleted and isospin
distillation becomes weaker with reduction in symmetry energy.
In panel (c), the isospin distribution for nuclei with $Z$=8 and 12 are
compared for both the symmetry energies at a density $\rho$
=0.02 fm$^{-3}$ for $X_t$=0.2. As expected, the distribution
gets wider with decrease in the symmetry energy. In panel (d),
the caloric curves at constant pressures $P$=0.002 and 0.02
MeV fm$^{-3}$ are displayed. The caloric curves are found
to be rather insensitive to the variation in the symmetry
energy coefficient.

\section{Concluding remarks}

 The equation of state of symmetric and asymmetric dilute 
nuclear matter and its various thermodynamic properties are
studied in the nuclear statistical equilibrium model and compared
with those obtained from the conventional mean-field model of
nuclear matter. The expressions for the pressure and density in the 
NSE model have the same structure as those obtained from the 
method of cluster expansion though the information content in the 
expansion coefficients is somewhat different. The clusterized 
matter is more stable than the bulk nuclear matter. The qualitative
behavior of the warm dilute matter towards compression or cooling
is the same in both the models. In the mean-field (NM) model,
the system separates itself in a bulk liquid and gas phase
at a transition density or a transition temperature; in the NSE
model, the system responds towards the changes by a marked
growth of clusters out of the dilute nucleonic gas at a transition
point. 

The isotherms in the two models look very similar. For symmetric 
nuclear matter, above the transition density, the pressure
remains constant in the density region we study. For asymmetric
matter, the pressure increases with density which becomes more
prominent with increasing asymmetry. This shows that like the
gas-liquid phase transition in the NM model \cite {mul}, the
transition to bulk clusterization in the NSE model 
for asymmetric matter occurs over
a temperature interval implying a continuous transition as 
opposed to that at constant temperature in the symmetric matter where
the transition is first order. This is further manifested in the
finite width in $c_p$ or a continuity in  
$S/A$ for asymmetric nuclear matter whereas for symmetric
matter, this is marked by a singularity in 
$c_p$ or a discontinuity in entropy.

In consonance with the recent experimental indications that the
nuclear symmetry energy coefficient of excited fragments produced
in a hot nuclear environment gets considerably reduced, calculations
have been performed to test the sensitivity of the nuclear EoS
with a weakened symmetry energy. Reduction in the symmetry
energy induces quantitative changes in both the inclusive and
exclusive nuclear observables, but qualitatively their behavior
remains similar; overall, an asymmetric matter with weakened
symmetry energy coefficient behaves as a regular system 
of effectively lesser asymmetry.

\acknowledgments{
The authors thank S. Mallick for fruitful discussions.
J.N.D. and S.K.S. gratefully acknowledge the financial support from
DST and CSIR, Government of India, respectively.}

\newpage
\centerline
{\bf Figure Captions}
\begin{itemize}
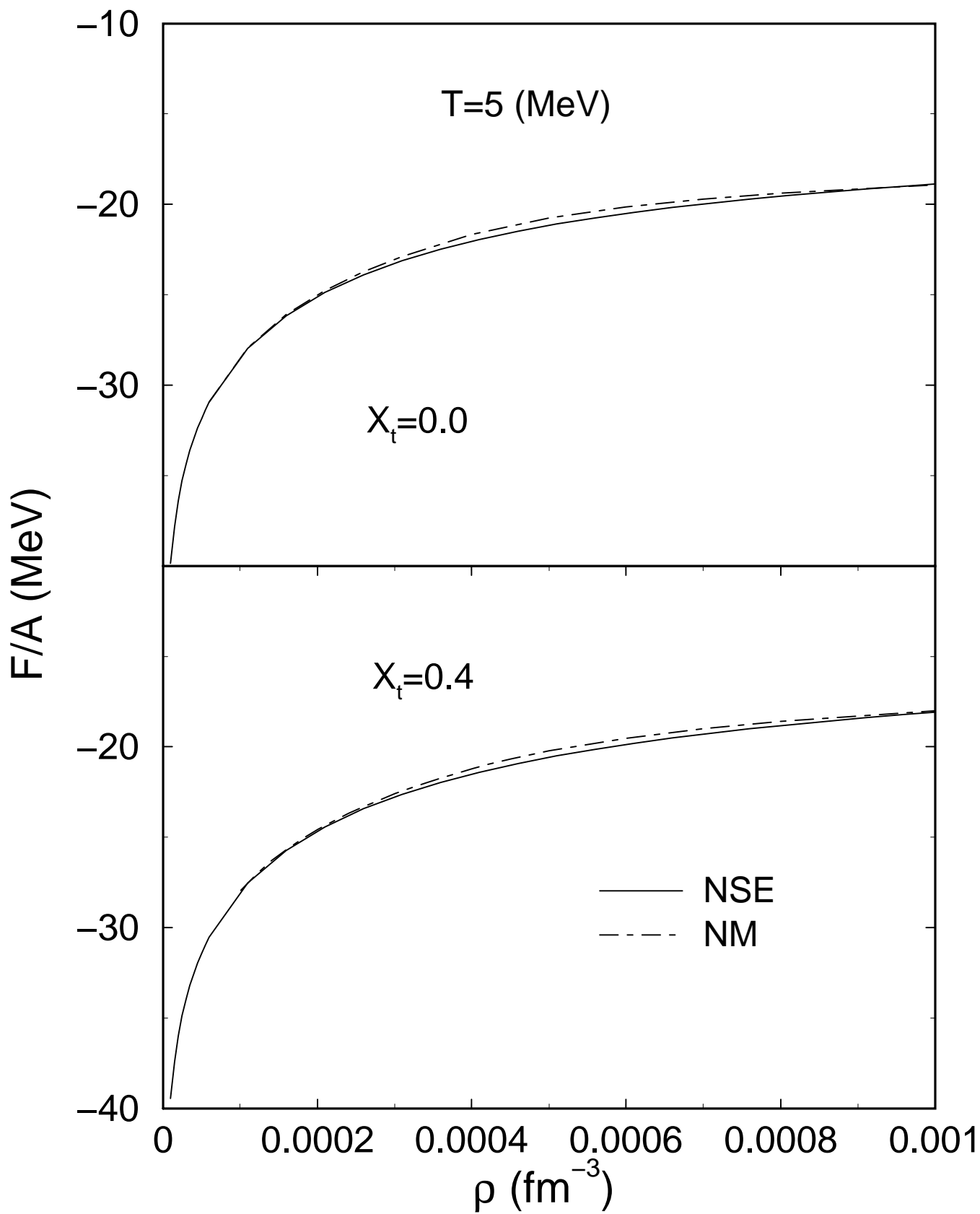
\item[Fig.\ 1] The free energy per nucleon shown  as a function of
density  for symmetric (upper panel) and asymmetric (lower panel)
nuclear matter at a temperature $T=$ 5 MeV. The full and dash lines
refer to the NSE and NM models, respectively.

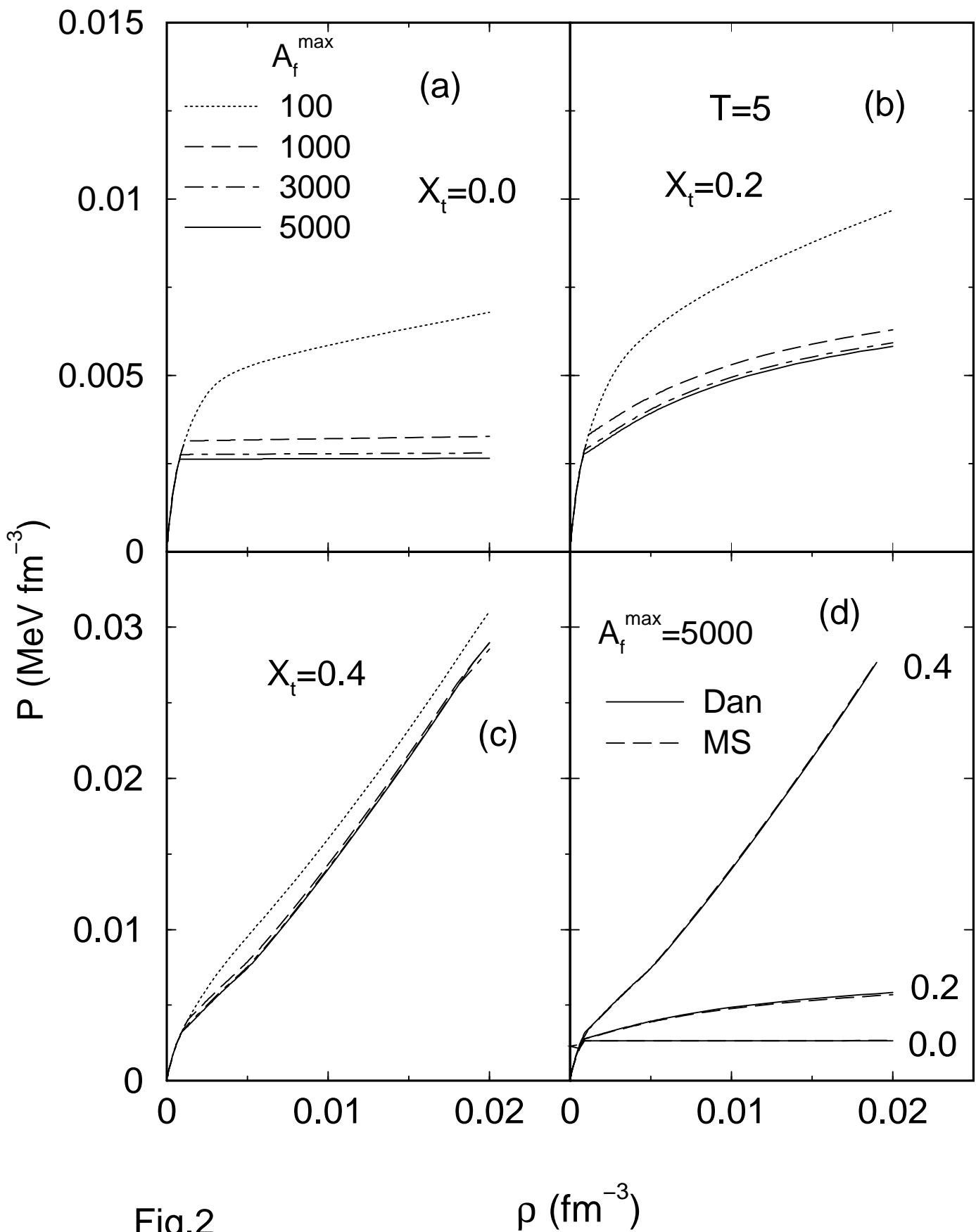
\item[Fig.\ 2] The sensitivity of the isotherm on the choice of
$A_f^{max}$ (see text) shown at $T=$ 5 MeV for symmetric and
asymmetric nuclear matter in the panels $(a), (b)$ and $(c)$.
Panel $(d)$ shows the dependence of isotherm on asymmetry with
Danielewicz (Dan) and Myers-Swiatecki (MS) mass formula.
  
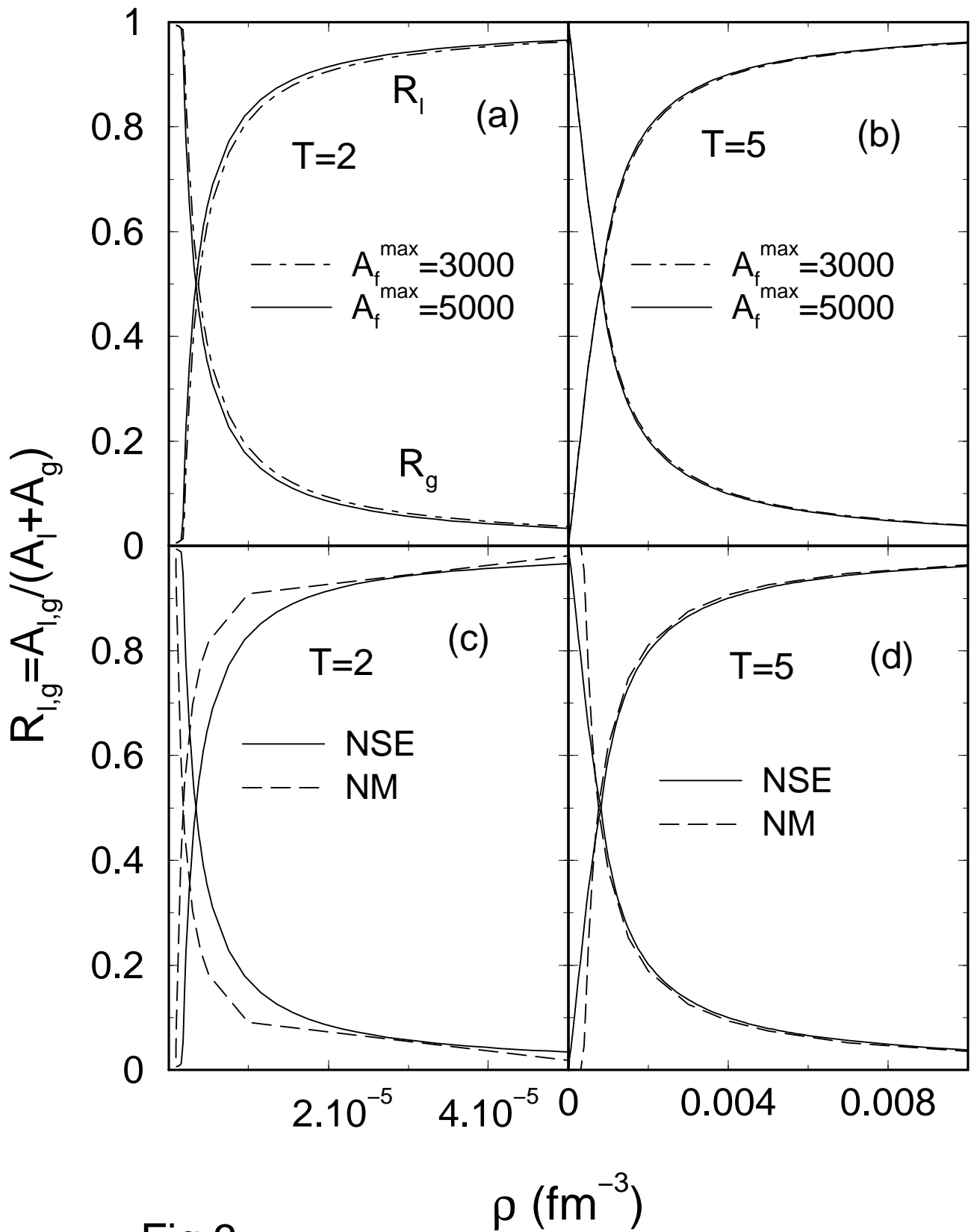
\item[Fig.\ 3] Dependence of nucleon fraction as a function of density in
the liquid ($R_l$) and in the gas $(R_g)$ phase on 
$A_f^{max}$ and temperature displayed in the panels $(a)$ and $(b)$
in the NSE model. The model dependence of the nucleon fractions
are shown in the panels $(c)$ and $(d)$ at two temperatures.
All the calculations are for symmetric nuclear matter.

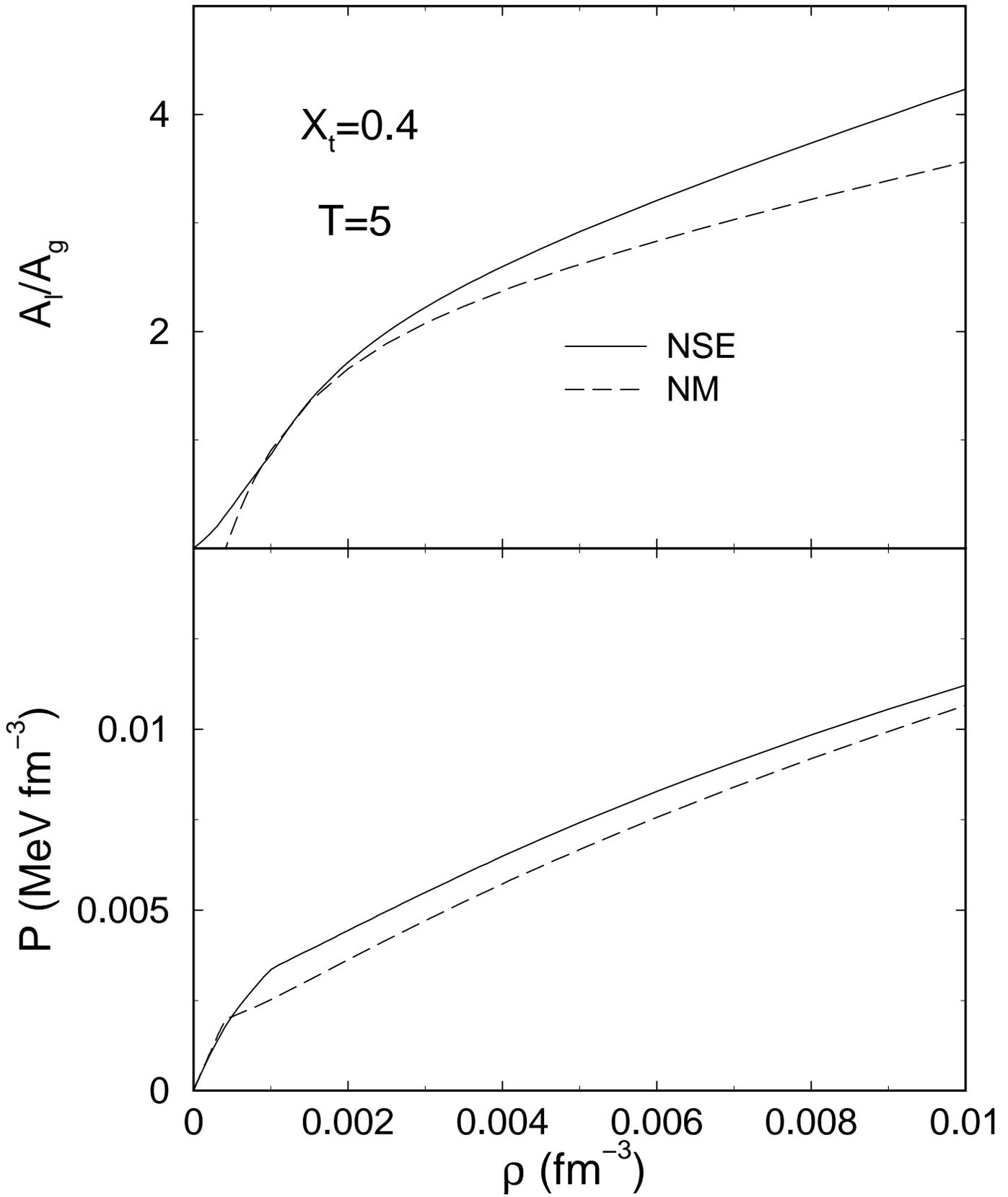
\item[Fig.\ 4] In the upper panel, the ratio of nucleons in the
liquid to that in the gas phase as a function of density 
at $T=$ 5 MeV is compared in the NM and NSE models
for asymmetric nuclear matter ($X_t=$0.4). The lower panel shows the
comparison of the isotherms for the same system in the two models.
  
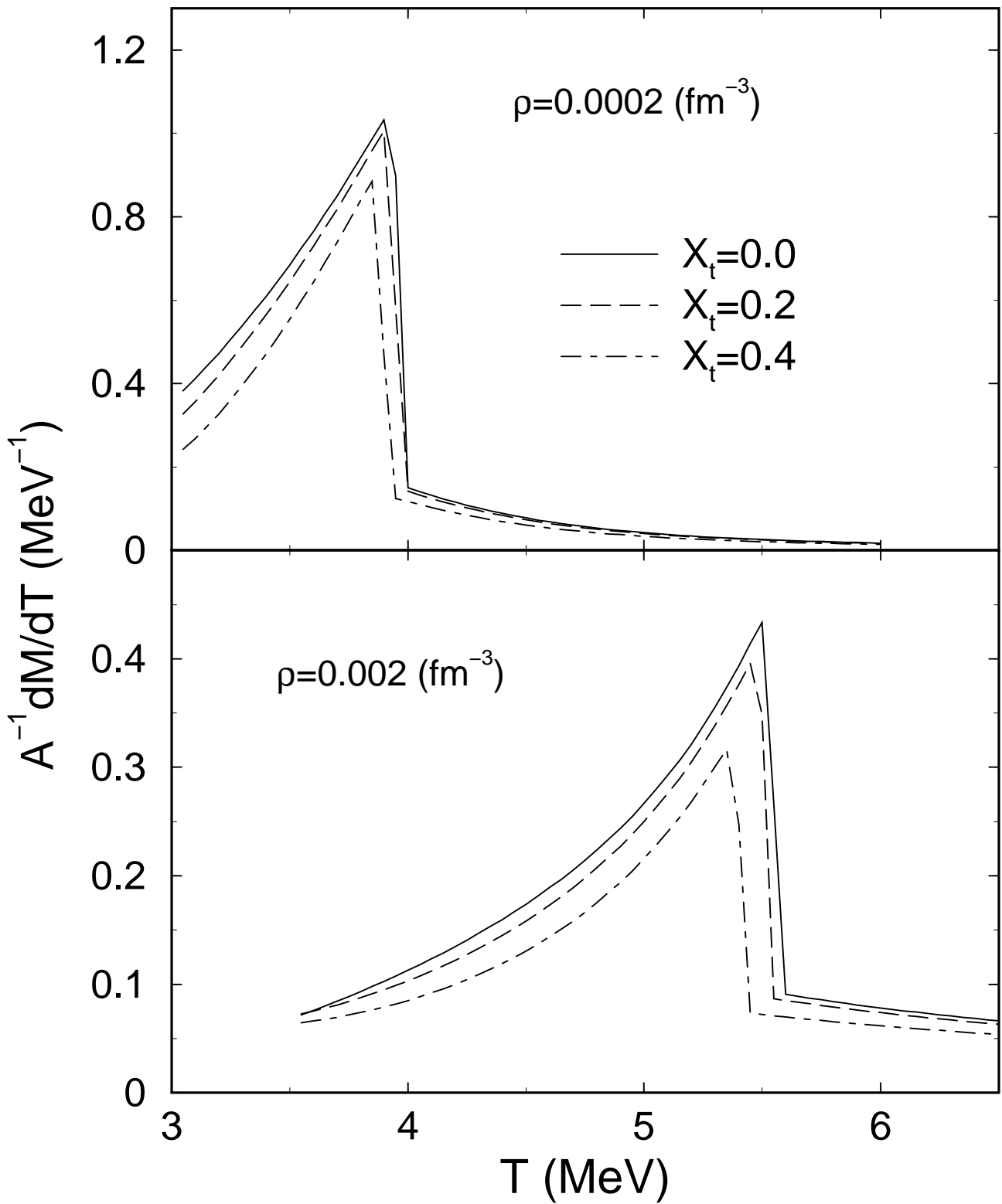
\item[Fig.\ 5] Multiplicity growth in isochoric cooling shown at
two densities for different asymmetries. 
  
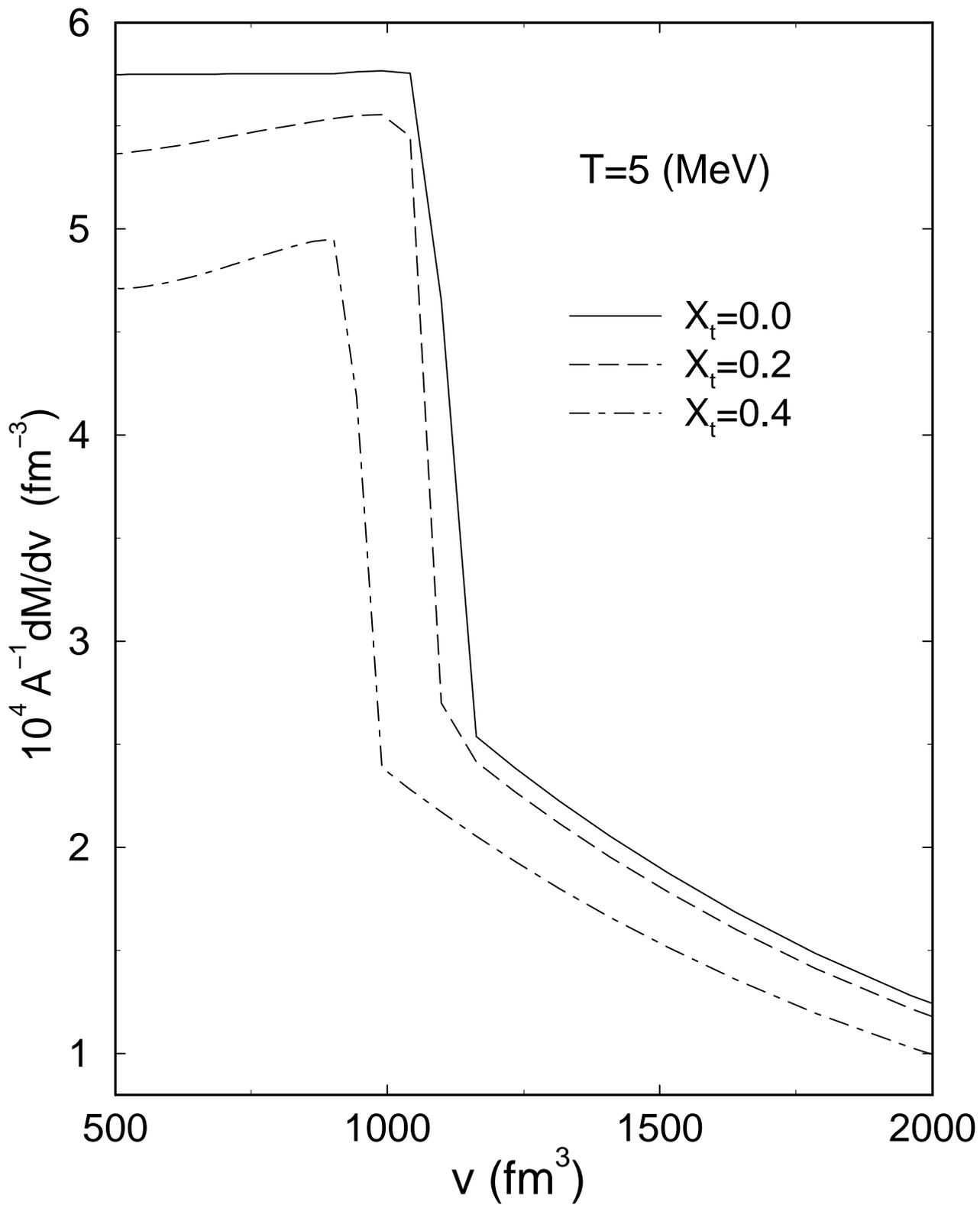
\item[Fig.\ 6] Multiplicity growth in isothermal compression at
$T=$ 5 MeV shown for different asymmetries.
  
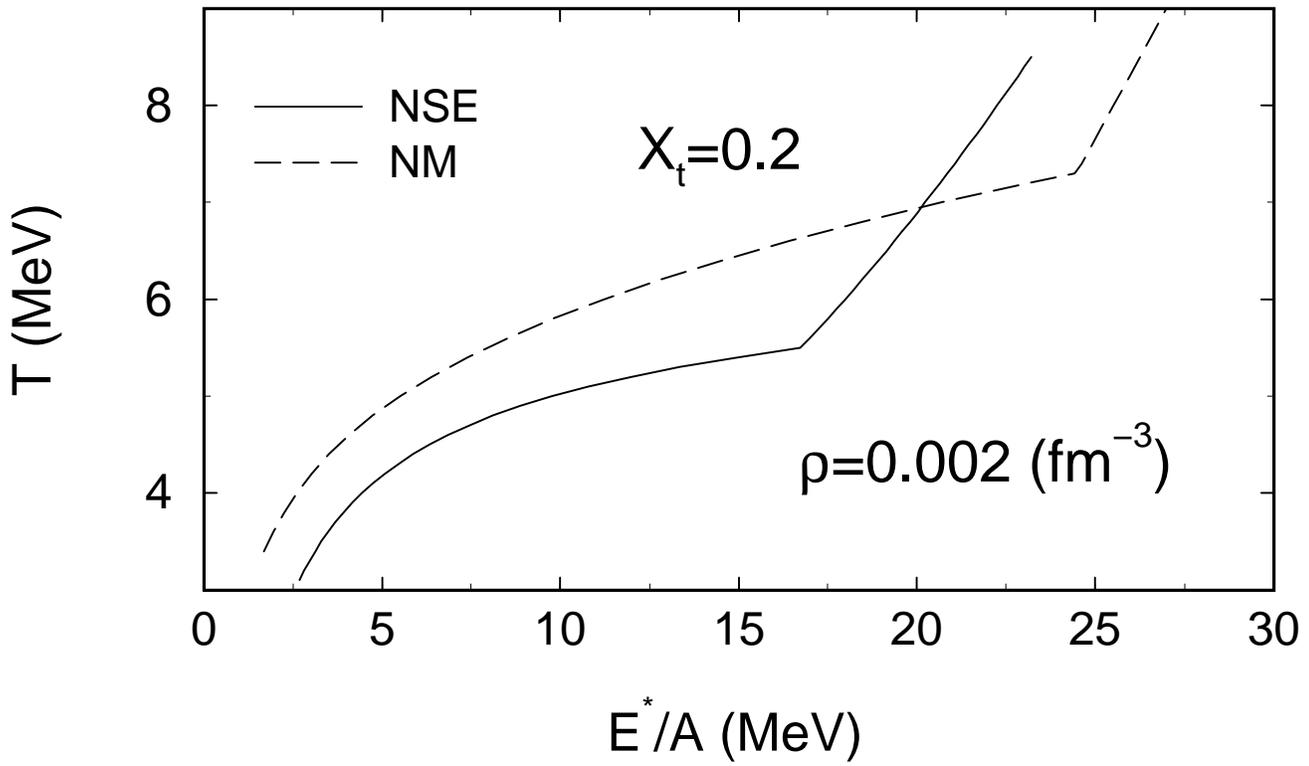
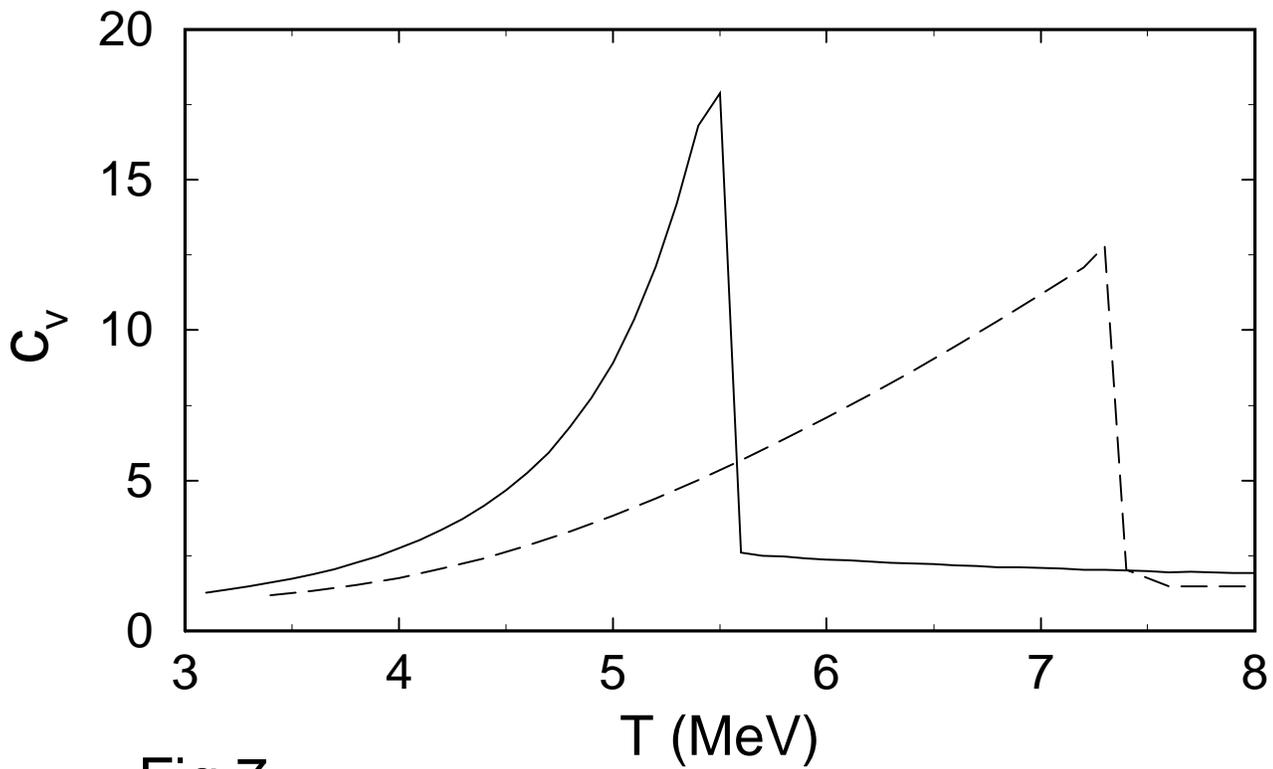
\item[Fig.\ 7] In the upper panel, caloric curves at constant
volume in the NM and NSE models for asymmetric nuclear matter 
with $X_t=$0.2 are shown. The corresponding specific heat per nucleon 
$c_v$ is shown in the lower panel.
  
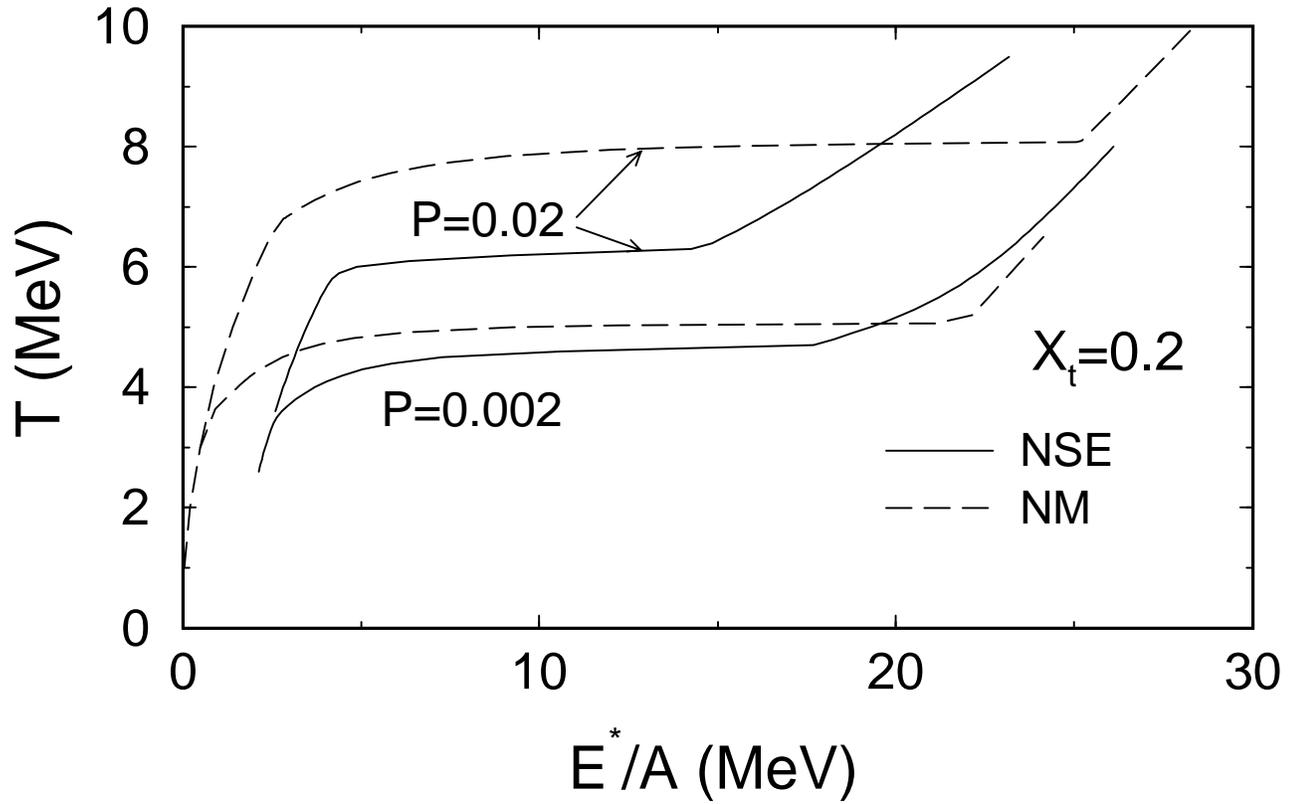
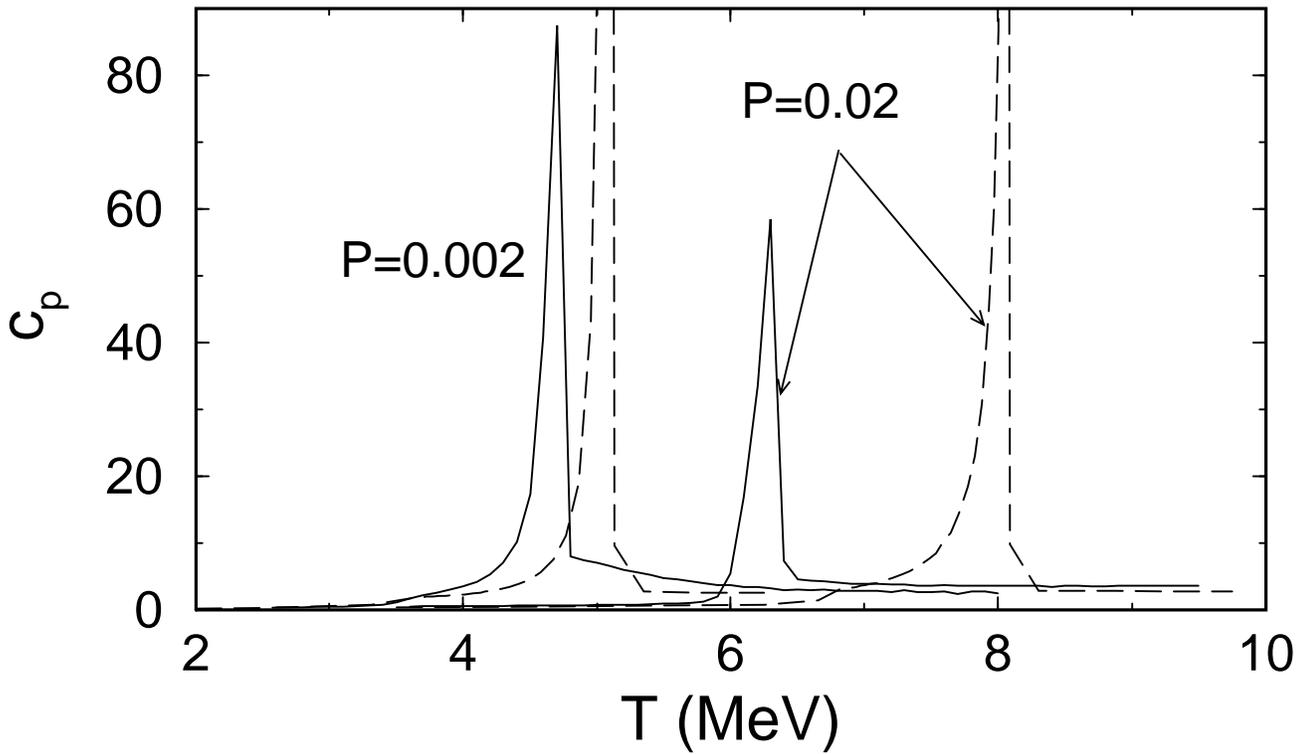
\item[Fig.\ 8] Caloric curves at two different pressures as
marked are shown in the NM and NSE models for asymmetric nuclear matter
($X_t=$0.2). The corresponding heat capacity $c_p$ 
per nucleon is shown in
the lower panel. 
 
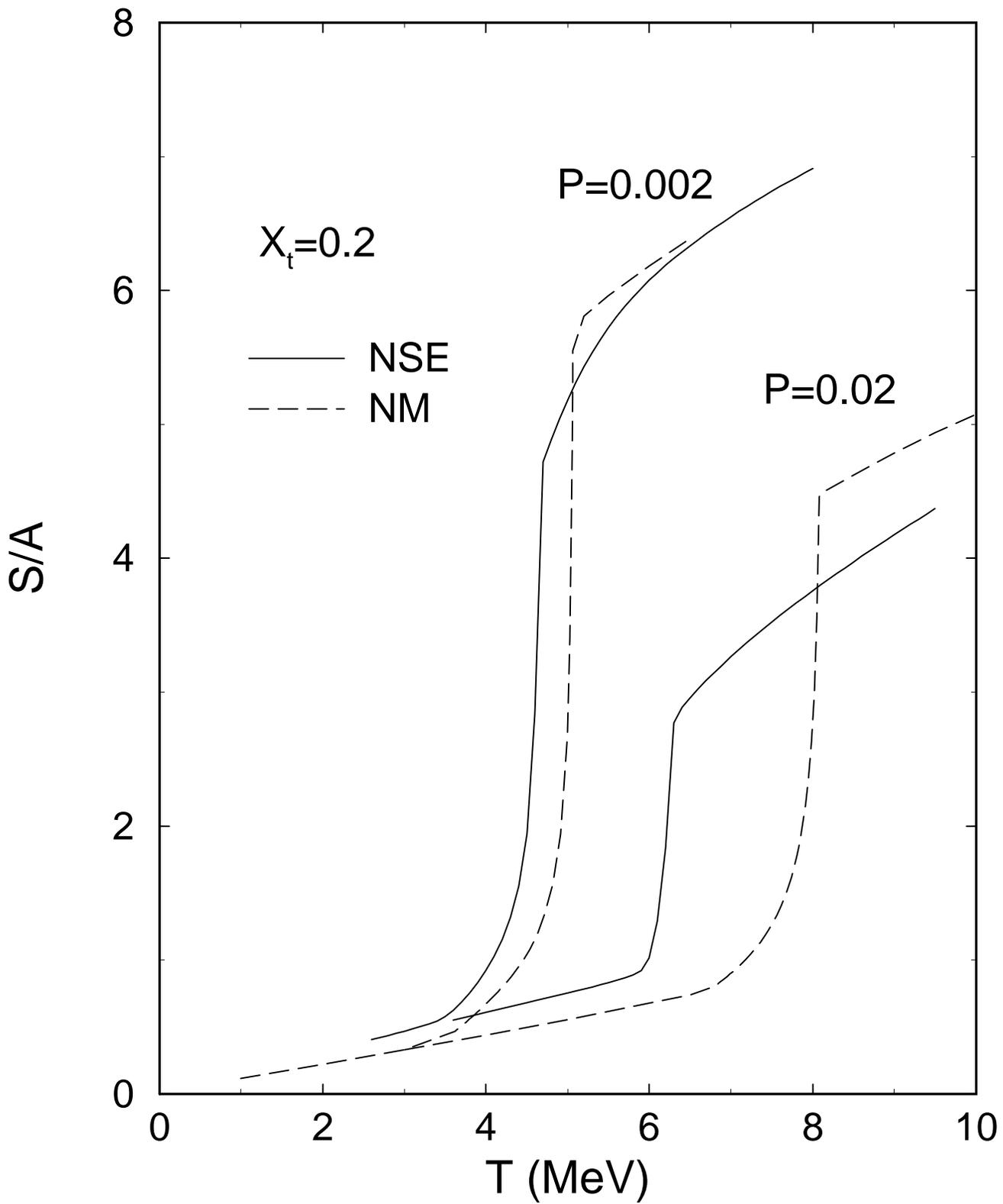
\item[Fig.\ 9] Evolution of entropy per nucleon with temperature
at two different pressures as marked calculated in the NM and NSE models
for asymmetric nuclear matter with $X_t=$0.2. 
  
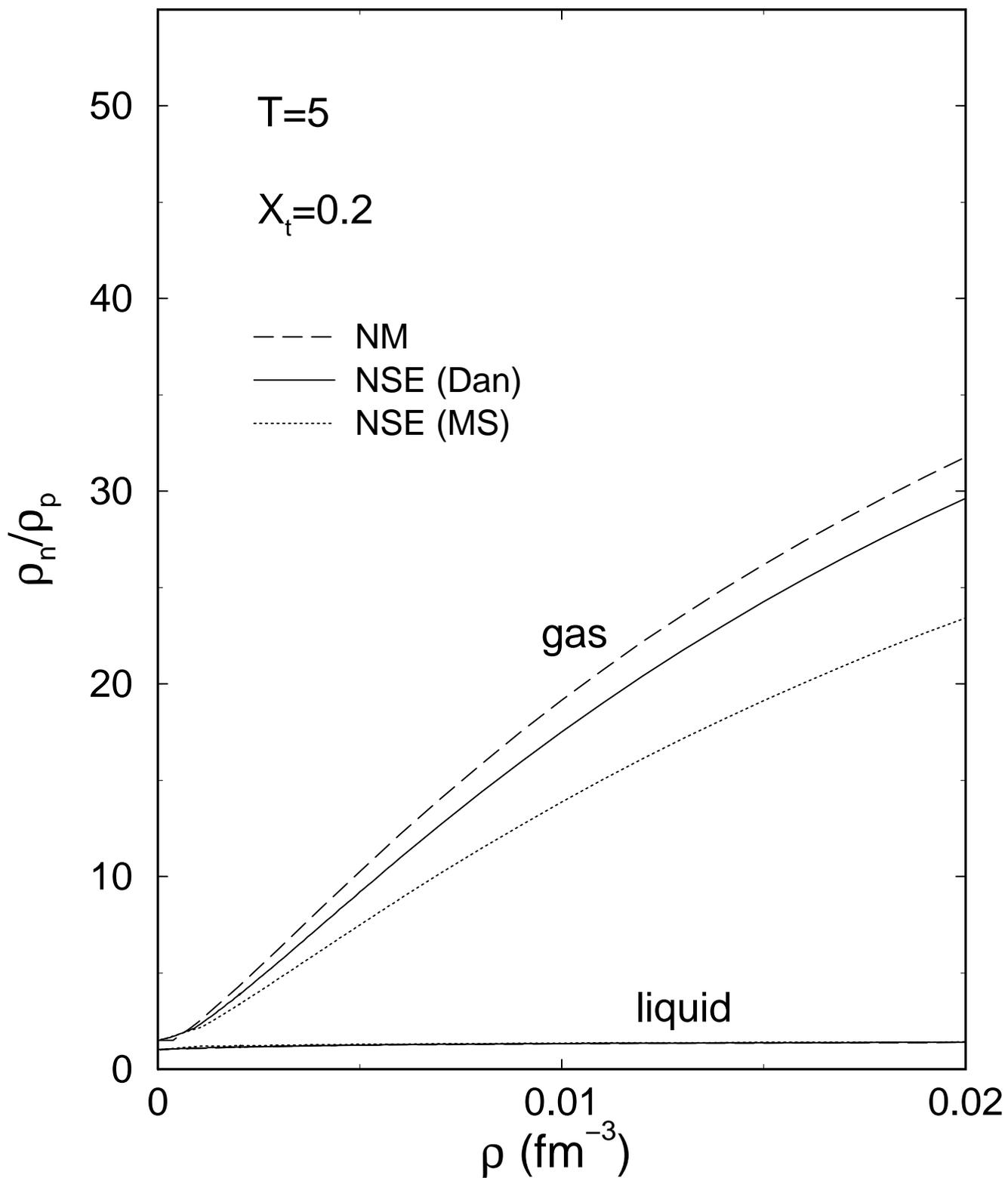
\item[Fig.\ 10] Evolution of neutron to proton density ratio
with isothermal compression at $T=$ 5 MeV in the liquid and gas 
phases for asymmetric nuclear matter ($X_t=$ 0.2) calculated in
the two models. Calculations in the NSE model have been performed
with both Danielewicz (Dan) and Myers-Swiatecki (MS) mass formulas.

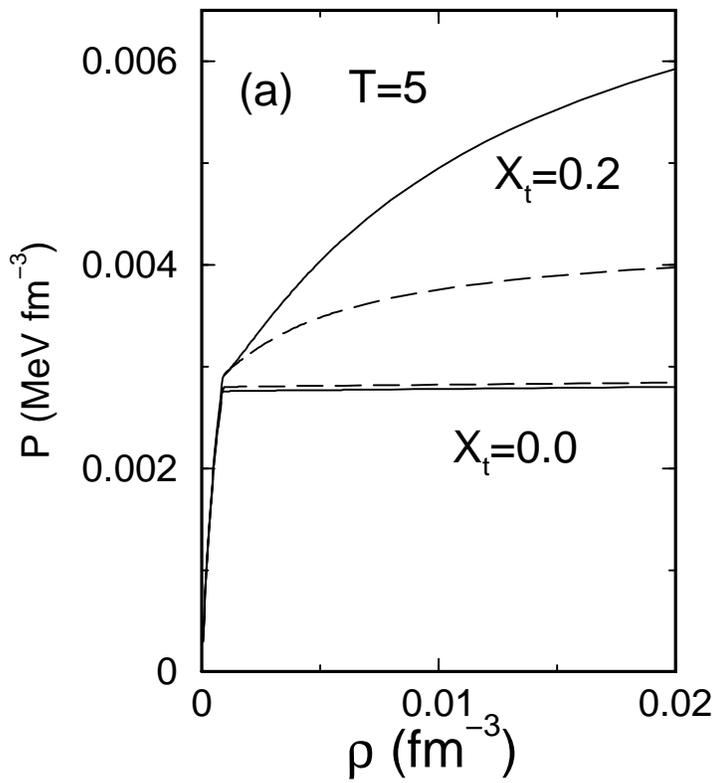
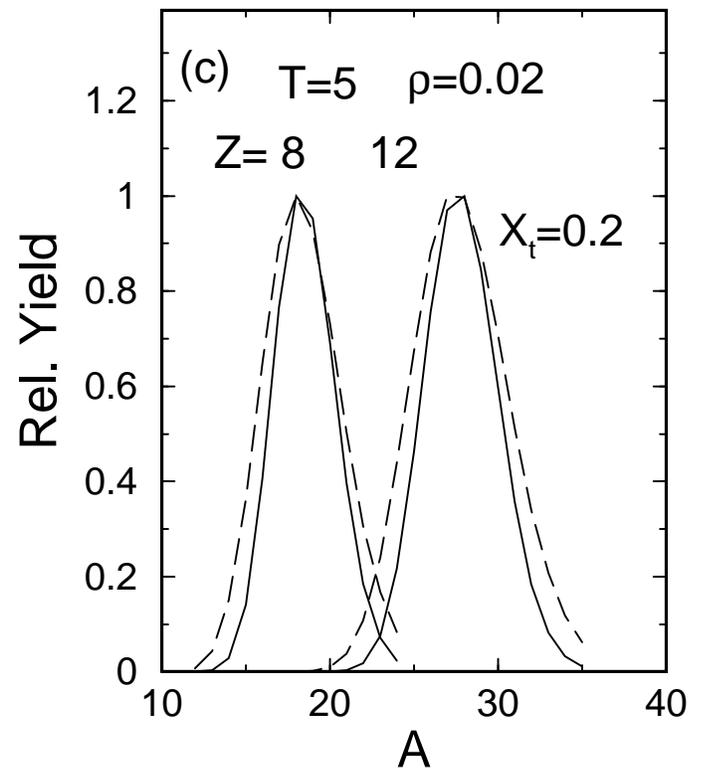
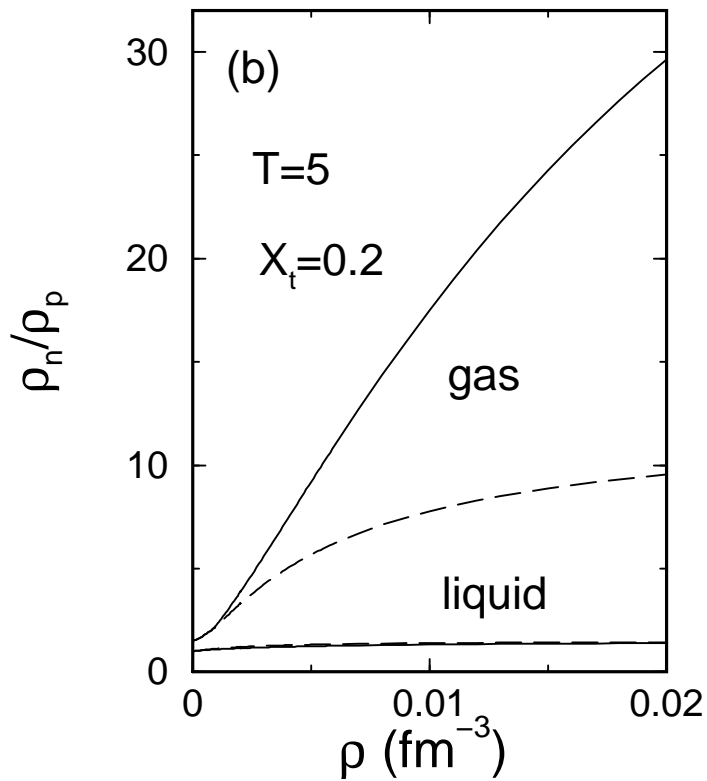
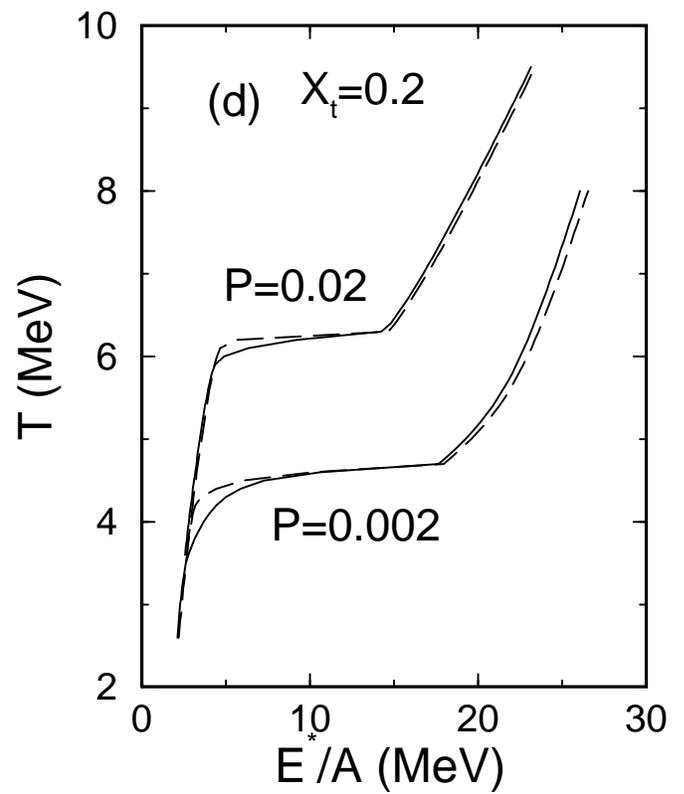
\item[Fig.\ 11] The effect of symmetry energy on a few observables
at $T$=5 MeV. The full and dashed lines correspond to calculations
with the original ground state value and a value reduced by 40$\%$
for the symmetry energy in Danielewicz mass formula, respectively.
The isotherms, neutron to proton ratio in the liquid and gas phases,
isotopic distributions and caloric curves at constant pressure
are shown in panels (a)-(d), respectively. For details, see text.

\end{itemize}

\end{document}